\documentclass[useAMS,usenatbib,longnamesfirst,usegraphicx]{mn2e}
\usepackage{txfonts}
\usepackage{color}
\usepackage{ulem}

\newcommand{\etal}{et al.}

\begin{document}
\title[Self-similarity of cooling superfluid neutron stars]{
Self-similarity relations for cooling superfluid neutron stars}

\author[P. S. Shternin and D. G. Yakovlev]{
P. S. Shternin$^{1,2}$\thanks{E-mail: pshternin@gmail.com},
D. G. Yakovlev$^{1}$
\\
$^{1}$Ioffe Institute,
Politekhnicheskaya 26, 194021 St.\ Petersburg, Russia \\
$^{2}$St.\ Petersburg Polytechnic University, Politechnicheskaya
29, 195251, St.\ Petersburg, Russia}

\date{Accepted . Received ; in original form}
\pagerange{\pageref{firstpage}--\pageref{lastpage}} \pubyear{2014}
\maketitle \label{firstpage}

\begin{abstract}
We consider models of cooling neutron stars with nucleon cores
which possess moderately strong triplet-state superfluidity of
neutrons. When the internal temperature drops below the maximum of
the critical temperature over the core, $T_{\rm C}$, this
superfluidity sets in. It produces a neutrino outburst due to
Cooper pairing of neutrons which greatly accelerates the cooling.
We show that the cooling of the star with internal temperature $T$
within $0.6\,T_{\rm C} \lesssim T \leq T_{\rm C}$ is described by
analytic self-similar relations. A measurement of the effective
surface temperature of the star and its decline, supplemented by
assumptions on star's mass, radius and composition of
heat-blanketing envelope, allows one to construct a family of
cooling models parametrized by the value of $T_{\rm C}$. Each
model reconstructs cooling history of the star including its
neutrino emission level before neutron superfluidity onset and the
intensity of Cooper pairing neutrinos. The results are applied to
interpret the observations of the neutron star in the Cassiopeia A
supernova remnant.
\end{abstract}


\section{Introduction} \label{sec:intro}

It is well known that observations of cooling neutron stars allow
one to explore still uncertain properties of superdense matter in
neutron star interiors \citep*[e.g.,][]{yakovlevpethick04}. Here,
we consider cooling of neutron stars with nucleon cores. We assume
that a star is not too young (with the age $t \gtrsim 10-200$ yr)
so that it is thermally relaxed (isothermal) inside except for the
thin heat-blanketing layer near the surface. In addition, we
assume that the star is on the neutrino cooling stage ($t \lesssim
10^5-10^6$ yr) meaning that it cools from inside via neutrino
emission from its interior (mainly from the core). The thermal
photon surface luminosity is much lower than the neutrino
luminosity and adjusts itself to the current internal thermal
state. In this way, the thermal emission from the surface reflects
the intensity of the neutrino emission that depends on the
properties of superdense matter in neutron star core.

There are two basic phenomena that can be tested by neutron star
cooling: (i) the operation of powerful direct Urca process of
neutrino emission in inner cores of massive neutron stars; (ii)
the presence of nucleon superfluidity in neutron star cores. The
direct Urca process is regulated by the symmetry energy of neutron
star matter (becomes allowed at sufficiently large symmetry energy
which results in rather large fraction of protons). To simplify
our analysis, we assume that the direct Urca process is not
allowed and focus on the effects of superfluidity. This is
equivalent of using the minimal cooling theory
\citep*{pageetal04,gusakovetal04}.

Following the standard minimal cooling theory we consider two
superfluids in the neutron star core -- singlet-state pairing of
protons and triplet-state pairing of neutrons. The appropriate
critical temperatures depend on the density $\rho$ and will be
denoted as $T_{\rm cp}(\rho)$ and $T_{\rm cn}(\rho)$,
respectively. Unfortunately, nucleon superfluidity is a very model
dependent phenomenon. Typically, the $T_{\rm cp}(\rho)$ and
$T_{\rm cn}(\rho)$ profiles over the stellar core have bell-like
shapes \citep*[e.g.,][]{lombardoschulze01}. Various models predict
very different profiles, so that it is instructive to consider
these profiles as unknowns and try to constrain them from
observations of cooling neutron stars.  As widely discussed in the
literature, the effects of proton and neutron superfluidities on
neutron star cooling are different \citep*[e.g.,][]{pageetal09}.
Proton superfluidity mainly suppresses neutrino emission processes
involving protons. As for neutron superfluidity, it also
suppresses the traditional processes of neutrino emission, but its
onset may initiate a powerful neutrino outburst due to Cooper
pairing of neutrons.  Neutron superfluidity occurs when the
temperature $T$ in the cooling star falls down to the maximum
critical temperature of neutrons in the core,
\begin{equation}
T_{\rm C}={\rm max} \{ T_{\rm cn}(\rho)\},
\label{e:TC}
\end{equation}
and can strongly accelerate the cooling.  This effect has been
used by \cite*{pageetal2011} and \cite*{shterninetal2011} to
interpret  the results by \cite{hoheinke09} and \cite{heinkeho10}
who analysed the observations of the neutron star in the
Cassiopeia A (Cas A) supernova remnant.

Following \cite*{pageetal2011} and \cite*{shterninetal2011}, we
consider the cooling scenario in which proton superfluidity is
much stronger than neutron one. Then, proton superfluidity appears
at the early cooling stage and suppresses neutrino emission
processes involving protons and proton heat capacity. Subsequent
cooling history contains two stages, prior ($T\geq T_{\rm C}$) and
after ($T<T_{\rm C}$) the onset of neutron superfluidity. The
first stage represents a slow cooling of the star. Such a cooling
is described by simple analytic relations which allow one to
perform model-independent analysis of the slow neutrino cooling
rate \citep*{yakovlevetal11}. Here, we focus on the second stage,
$T<T_{\rm C}$, and show that as long as $T\gtrsim 0.6 T_{\rm C}$
neutron star cooling is described by self-similar analytic
equations which can be used to reconstruct the cooling history of
the star from observational data.

\section{Cooling equations}
\label{s:equations}

We follow the cooling theory of neutron stars with isothermal
interiors at the neutrino cooling stage
\citep[e.g.,][]{yakovlevetal11}. The basic cooling equation
(including the effects of General Relativity) is
\begin{equation}
   \frac{{\rm d} T}{{\rm d}t}=-\ell(T)=-\frac{L_\nu(T)}{C(T)}.
\label{e:cool}
\end{equation}
Here, $T$ is the {redshifted internal temperature}, $t$ is
Schwarzschild time, $\ell(T)$ is the neutrino cooling rate,
$L_\nu(T)$ is {the neutrino luminosity} and $C(T)$ is the
integrated heat capacity of the star. It is the redshifted
temperature $T$ which is constant over the isothermal internal
region of the star; $L_\nu(T)$ and $C(T)$ in equation
(\ref{e:cool}) also have to be redshifted. If $\ell(T)$ is known,
one can immediately write down a formal solution of the cooling
problem
\begin{equation}
     t-t_{\rm i}=\int_T^{T_{\rm i}} \frac{{\rm d}T'}{\ell(T')},
\label{e:coolint}
\end{equation}
where $T_{\rm i}$ is the temperature at some initial moment of
time $t=t_{\rm i}$. This expression describes the evolution of the
internal temperature $T(t)$. The surface temperature of the star,
$T_{\rm s}(t)$, can be calculated then from the internal one using
the relation between the internal and surface temperatures
\citep*[e.g.,][]{potekhinetal97}.

Let $t=t_{\rm C}$ refer to the onset of neutron superfluidity in
the neutron star core (at $T=T_{\rm C}$, where $T_{\rm C}$ has to
be treated as the maximum value of the redshifted critical
temperature for neutron superfluidity in the stellar core). Before
the onset, we have a slow cooling with $L_\nu(T) \propto T^8$,
$C(T) \propto T$ and $\ell(T)\propto T^7$. In our notations, this
slow cooling is described by
\begin{equation}
      t=\frac{t_{\rm C}}{\tau^6}, \quad \mathrm{at}~\tau \equiv \frac{T}{T_{\rm C}}
            \geq 1
            ~~{\rm that~ is}~~t \leq t_{\rm C}.
\label{e:slowcool}
\end{equation}
Note that in this case
\begin{equation}
   \ell(T)=\frac{T}{6t}.
\label{e:lslowcool}
\end{equation}
This solution is obtained with the standard initial condition
widely used in the  neutron star cooling problem: $T_{\rm i} \to
\infty$ as $t_{\rm i} \to 0$. Here and below, we consider the
cooling solutions $t=t(\tau)$ as functions of the dimensionless
quantity $\tau$.

After the neutron superfluidity onset, from equation (\ref{e:coolint}) we have
\begin{equation}
    t=t_{\rm C}+\int_T^{T_{\rm C}} \frac{{\rm d}T'}{\ell(T')} \qquad \mathrm{at}~~t>t_{\rm C}.
\label{e:cool-SF}
\end{equation}

At this stage, we need the neutrino cooling rate
$\ell(T)=\ell_0(T)+\ell_{\rm CP}(T)$ which includes the slow
neutrino cooling [$\ell_0(T)=\ell_{\rm C}\,\tau^7$, $\ell_{\rm
C}=\ell(T_{\rm C})$] and an extra cooling $\ell_{\rm
CP}(T)=L_\nu^{\rm CP}(T)/C(T)$ due to Cooper pairing of neutrons.

The Cooper pairing neutrino luminosity $L_\nu^{\rm CP}(T)$ has to
be calculated by integration of the appropriate neutrino
emissivity over the superfluid layer in the neutron star core
\citep{gusakovetal04}. When the star cools, the layer becomes
wider. The neutrino emissivity is a complicated function of $T$.
Generally, the temperature dependence $L_\nu^{\rm CP}(T)$ is
sensitive to the employed model of the star and to the model of
$T_{\rm cn}(\rho)$. However, as noticed by \citet{gusakovetal04},
as long as $T$ is not much lower than $T_{\rm C}$, a superfluid
layer of the core is not too wide. It is located in the vicinity
of the $T_{\rm cn}(\rho)$ peak, and the peak can be approximated
by an inverted parabolic function of radial coordinate $r$ within
the star centred at some $r=r_{\rm C}$. In this case, the
integrated neutrino luminosity $L_\nu^{\rm CP}(T)$ becomes a {\it
universal} function of $\tau$. It was calculated by
\citet{gusakovetal04} who approximated their result by an
analytical expression (their eqs. 5 and 7). Their expression is
cumbersome but we note that it is accurately described by a much
simpler formula
\begin{equation}
 \ell_{\rm CP}(T)=116 \,\delta \, \ell_{\rm C} \tau^7 (1-\tau)^2,
\label{e:lCP}
\end{equation}
where $\delta$ is {\it a convenient dimensionless parameter that
measures the efficiency of Cooper pairing neutrino emission} with
respect to the slow cooling level (see below). We expect that this
universal formula is valid at $0.6 \lesssim \tau\leq 1$ ($0.6
T_{\rm C} \lesssim T \leq T_{\rm C}$), although the factor 0.6 is
conditional [depends on $T_{\rm cn}(\rho)$ model]. Notice that at
$\tau$ very close to 1, our approximation (\ref{e:lCP}) is
slightly inaccurate because, actually, at these $\tau$ the
function $\ell_{\rm CP}(\tau)$ behaves as $(1-\tau)^{3/2}$.
Neglecting this effect allows us to obtain analytic solution of
the cooling problem at $\tau \gtrsim 0.6$.

Our approach tacitly assumes that neutron superfluidity does not
affect the heat capacity $C(T) \propto T$. This is generally not
true because the specific heat capacity of neutrons is affected by
superfluidity \citep*[see, e.g.,][]{yakovlevetal99}. However, at
the early superfluid stage, the superfluid layer within the
stellar core is not too wide, so that it contributes little to the
integrated heat capacity $C(T)$, and the assumption is justified.
Using the same arguments, we also neglect the reduction of the
slow component of the neutrino cooling rate, $\ell_0(T)$, by
neutron superfluidity.

\begin{figure}
\includegraphics[width=0.45\textwidth]{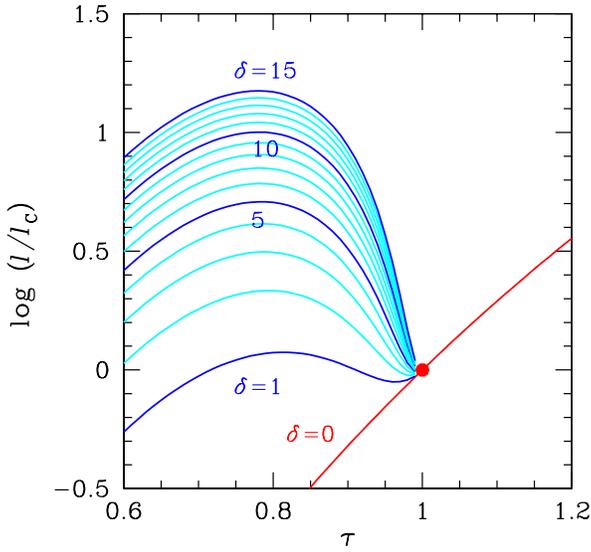}
\caption{(Color on line) Normalized neutrino cooling rate
$\ell(T)/\ell_{\rm C}$ versus $\tau=T/T_{\rm C}$ for $\delta=0$,
1,\ldots, 15. The $\delta=0$ curve is for a star without neutron
superfluidity in the core. In other cases, neutron superfluidity
of different efficiency $\delta$ sets in at $T=T_{\rm C}$ (filled
dot). This superfluidity intensifies neutrino cooling, with the
maximum of the extra Cooper pairing neutrino cooling rate
$\ell_{\rm CP}(T_{\rm m})$ at $T=T_{\rm m}=0.77\,T_{\rm C}$.
\label{fig:lvsT} }
\end{figure}

Under these assumptions, the total neutrino cooling rate at the
early superfluid cooling stage ($0.6\,T_{\rm C} \lesssim T \leq T_{\rm C}$) is
\begin{equation}
     \ell(T)=\ell_{\rm C}\,\tau^7 \left[ 1+116 \,\delta  \, (1-\tau)^2 \right].
\label{e:total-ell}
\end{equation}
The dependence of the neutrino cooling rate (\ref{e:total-ell}) on
$\tau$ for $\delta=0,1, \ldots, 15$ is plotted in Fig.\
\ref{fig:lvsT}. When the temperature decreases, $\ell(T)$ rapidly
decreases too as long as neutron superfluidity is absent (at $\tau
\geq 1$). After the superfluidity onset, $\ell(T)$ grows up
because the Cooper pairing neutrino emission starts to operate.
Then, it reaches maximum and decreases again as neutron
superfluidity becomes older. The maximum of the Cooper pairing
neutrino cooling rate $\ell_{\rm CP}(T)$ takes place at $T=T_{\rm
m}=0.77\, T_{\rm C}$ ($\tau_{\rm m}=0.77$). The maximum value of
this rate is $\ell_{\rm CP}(T_{\rm m})=\ell_{\rm C}\, \delta$.
Accordingly, $\delta$ is the ratio of two neutrino cooling rates,
\begin{equation}
 \delta=\ell_{\rm CP}(T_{\rm m})/\ell_{\rm C},
\label{e:def-delta}
\end{equation}
at $T=T_{\rm m}$ and $T=T_{\rm C}$. Note that $\ell_{\rm
CP}(T_{\rm m})/\ell_0(T_{\rm m})=5.8\,\delta$ and $\ell_{\rm
CP}(0.6\,T_{\rm C})/ \ell_0(0.6\,T_{\rm C})=18.56\,\delta$.

Recall that in our approach, the critical temperature $T_{\rm
cn}(\rho(r))$ as a function of radial coordinate within the
stellar core is approximated by an inverted parabola. It is
determined by two parameters -- the peak temperature $T_{\rm C}$
and a characteristic peak width $\delta r_{\rm C}$. Using the
results by \citet{gusakovetal04}, one can show that
\begin{equation}
    \delta =A(r_{\rm C})\, \delta r_{\rm C}/T_{\rm C},
\label{e:deltasim}
\end{equation}
where $A(r_{\rm C})$ is some function of $r_{\rm C}$  which is the
position of the maximum $T_{\rm cn}$ in the core. Therefore, if we
fix the neutron star model and the shape of $T_{\rm cn}(\rho)$
profile (i.e. $\delta r_{\rm C}$ and $r_{\rm C}$) but increase
$T_{\rm C}$, we would lower $\delta \propto 1/T_{\rm C}$ (i.e.,
lower the efficiency of the Cooper pairing neutrino cooling).

In the case of mature neutron superfluidity ($\tau\lesssim 0.2$,
not shown in Fig.~\ref{fig:lvsT}), the neutrino cooling rate is
not described by equation~(\ref{e:total-ell}) anymore. It can be
shown \citep{gusakovetal04, pageetal04}, that in this limit
neutrino cooling rate behaves as $\ell(T)\propto T^7$. Therefore,
the cooling of the star with mature superfluidity mimics the
standard cooling, equation~(\ref{e:slowcool}), but at higher
cooling rate.

%

\begin{figure}
\includegraphics[width=0.45\textwidth]{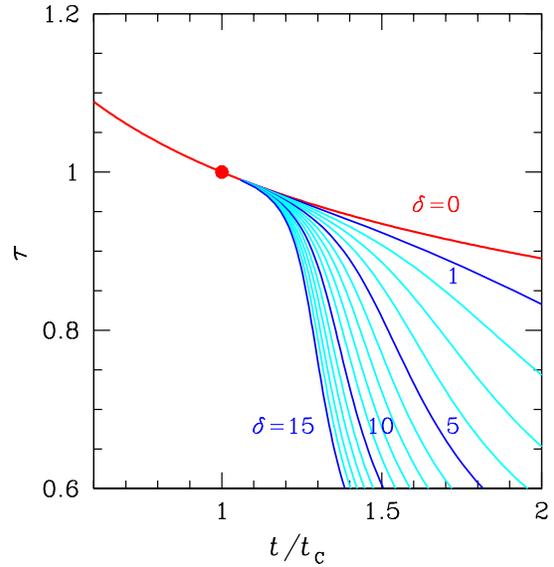}
\caption{(Color on line) Normalized cooling curves $\tau=T/T_{\rm
C}$ versus $t/t_{\rm C}$ for $\delta=0$, 1,\ldots, 15 (same  as in
Fig.\ \ref{fig:lvsT}). After the onset of neutron superfluidity
(at $t \geq t_{\rm C}$), the star is colder for larger $\delta$
due to stronger Cooper pairing neutrino emission.
\label{fig:Tvst} }
\end{figure}

\begin{figure}
\includegraphics[width=0.45\textwidth]{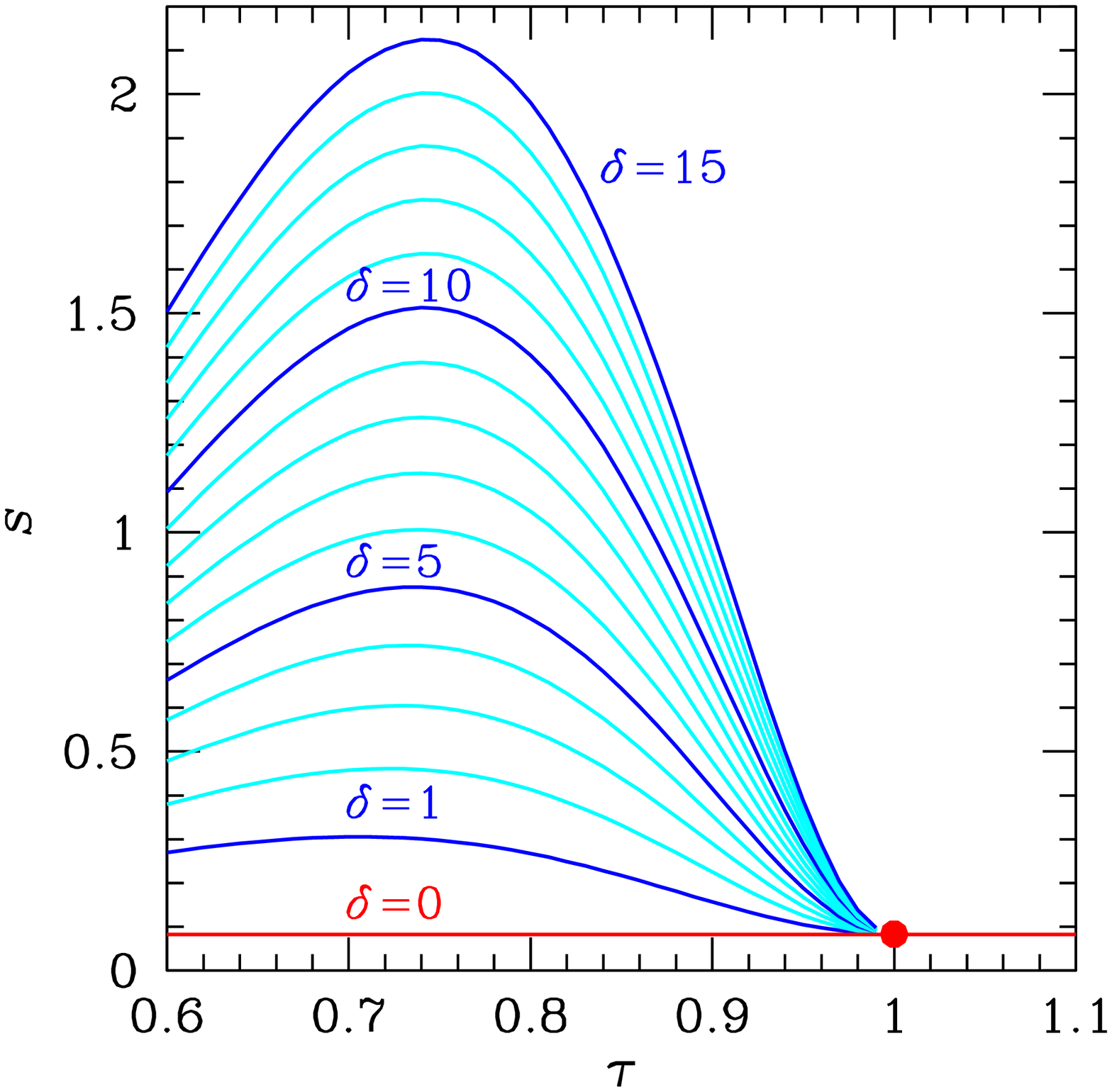}
\caption{(Color on line) The slope $s$ of the surface temperature
decline versus $\tau$ for the cooling solutions with $\delta=0$,
1,\ldots, 15 presented in Figs.\ \ref{fig:lvsT} and
\ref{fig:Tvst}. \label{fig:svsT} }
\end{figure}

\begin{figure}
\includegraphics[width=0.45\textwidth]{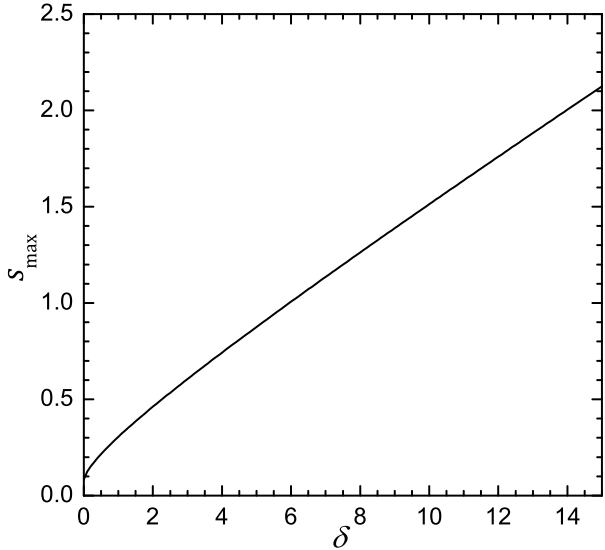}
\caption{The maximum value $s_{\rm max}$ of the surface temperature
decline versus $\delta$. \label{fig:sm} }
\end{figure}

Now, we substitute equation (\ref{e:total-ell}) into equation
(\ref{e:cool-SF}). The integral is taken analytically, and we
obtain
\begin{equation}
t=t_{\rm C}\,\left[1 +6 \,I_7(\tau)\right] \qquad \mathrm{at}~~t>t_{\rm C},
\label{e:cool-SF1}
\end{equation}
where $I_7(\tau)$ belongs to a family of integrals
\begin{equation}
  I_m(\tau)=\int_\tau^1 \frac{{\rm d}x}{x^m\,[1+116\;\delta\; (1-x)^2]},
\label{e:Im}
\end{equation}
with integer $m=0,1,\dots$. These integrals are presented in
 Appendix \ref{s:app_im}.

Now, any cooling solution can be easily calculated using equations
(\ref{e:slowcool}) and (\ref{e:cool-SF1}) for given $\delta$ and
$\tau \gtrsim 0.6$. It is clear that such solutions are
selfsimilar. Appropriate thermal evolution of a star is easily
understood from Figs.\ \ref{fig:lvsT}--\ref{fig:sm}. We have
already described Fig.\ \ref{fig:lvsT} which shows $\ell(T)$ for
$\delta=0$,\ldots, 15. Fig.\ \ref{fig:Tvst} presents cooling
curves (the internal temperature $T$ versus age $t$ in
dimensionless units for the same values of $\delta$). The higher
$\delta$ (the efficiency of Cooper pairing neutrino cooling), the
cooler the star after superfluidity onset. Fig.\ \ref{fig:svsT}
shows the evolution of potentially important observable
\begin{equation}
  s=-\frac{{\rm d}\,\ln T_{\rm s}(t)}{{\rm d}\,\ln t},
\label{e:defs}
\end{equation}
which is the (minus) logarithmic time derivative of the effective
surface temperature $T_{\rm s}$ of the star [in other words, the
slope of the cooling curve, $T_{\rm s}(t)$]. This dimensionless
quantity can be measured if cooling of a neutron star is observed
in real time. In order to calculate $s$, we need to relate the
internal temperature $T$ to $T_{\rm s}$. It is well known (e.g.,
Potekhin et al. 1997) that the $T_{\rm s}$--$T$ relation depends
on the composition of the outer heat-blanketing envelope of the
star (because the composition affects thermal conductivity within
the envelope). However, this relation is well approximated by a
power law, $T_{\rm s} \propto T^\beta$, with $\beta \approx 0.5$.
This makes $s$ (but not $T_{\rm s}$ itself!) {\it almost
insensitive to the composition of the envelope}. The curves in
Fig.\ \ref{fig:svsT} are calculated assuming $\beta=0.5$ from the
equation
\begin{equation}
   s=\beta \; \frac{{\rm d}\,\ln T}{{\rm d}\,\ln t}=\frac{\beta }{6\tau}\,
    \frac{t}{t_{\rm C}}\,\frac{\ell(T)}{\ell_{\rm C}}.
\label{e:s-calc}
\end{equation}

If neutron superfluidity is absent in the core ($\delta=0$ in
Fig.\ \ref{fig:svsT}) and cooling is slow, we obtain $s=s_0=1/12$.
However, soon after the neutron superfluidity onset during the
Cooper pairing neutrino outburst, $s$ strongly increases, reaches
maximum and then decreases again to $s_0=1/12$ after superfluidity
develops in the core  (not shown in Fig.~\ref{fig:svsT}). The
enhanced values of $s(T)$ trace the neutrino cooling function
$\ell(T)$ and can serve as a sensitive indicator of a neutrino
outburst in the neutron star core. Because of the peak behaviour
of $s$ as a function of $T$ or $t$, there is a maximum value
$s=s_{\rm max}(\delta)$ for any given solution; it is reached near
the maximum of the neutrino outburst. Therefore, any value of $s$
within the range $1/12 \leq s \leq s_{\rm max}$ is realized twice,
before and after the maximum. The dependence of $s_{\rm max}$ on
$\delta$ is plotted in Fig.~\ref{fig:sm}. Equivalently, the figure
presents the minimal value of $\delta$ required to reach a given
value of $s$.

\section{Data analysis using cooling solutions}
\label{s:analysis}

\subsection{The Cas A neutron star}
\label{s:CasA}

{\citet{yakovlevetal11} have found that the cooling of neutron
stars regulated by the modified Urca process of neutrino emission
is fairly independent of the equation of state (EOS) of dense
stellar cores so that such stars can be used as standard cooling
candles. This allowed the authors to develop a simple procedure
for a model-independent analysis of the neutrino emission rates of
slowly cooling stars in terms of standard candles (see also Sec.\
3.3). This procedure is independent of the EOS and particular
processes of slow neutrino emission in the core. Here, we extend
it to the case when the early slow cooling is accelerated by the
neutron superfluidity onset.}

By way of illustration, consider the neutron star with the carbon
atmosphere in the Cas~A supernova remnant, which is currently the
only isolated neutron star whose cooling in real time is possibly
observed \citep{heinkeho10}. Note that after the first explanation
of this effect by \citet{pageetal2011} and
\citet{shterninetal2011}, several alternative explanations have
been proposed (e.g, \citealt*{yangetal2011},
\citealt*{negreirosetal2013}, \citealt*{sedrakian2013},
\citealt*{blaschkeetal2013}, \citealt*{bonannoetal2014}).
Moreover, the presence of real-time cooling itself has been put
into question by \citet{posseltetal2013} who attribute it to the
{\it Chandra} ACIS-S detector degradation in soft channels. More
observations are needed to resolve this issue.

A detailed analysis of the Cas~A surface temperature decline has
been done recently by \citet{elshamouty13} by comparing the
results from all the {\it Chandra} detectors. They find the
weighted mean of the decline rate as $2.9\%\pm0.5_{\rm stat}\%\pm
1_{\rm sys}\%$ over the 10 yr base using information from all
detectors, and $1.4\%\pm0.6_{\rm stat}\%\pm 1_{\rm sys}\%$
excluding the data from the ACIS-S detector in the graded mode
which can suffer from the grade migration \citep{elshamouty13}.
With the age of the Cas~A supernova remnant and its  central
neutron star $t_{\rm d} \approx 330$~yr, this corresponds (in our
notations) to the current ($t=t_{\rm d}$) values $s=s_{\rm
d}=0.96\pm0.16_{\rm stat}\pm0.33_{\rm sys}$ and $0.46\pm 0.20_{\rm
stat}\pm 0.33_{\rm sys}$, with and without ACIS-S(G) data,
respectively. Recall that the standard slow cooling requires
$s=1/12\approx 0.08$. The measured effective surface temperature
(non-redshifted to a distant observer) is $T_{\rm s}\approx 2$~MK
(\citealt{hoheinke09}, see also \citealt{yakovlevetal11}). The
spectral fits do not constrain the mass and radius of the neutron
star.  To be specific, we select particular $M=1.65\,{\rm
M_\odot}$ and $R=11.8$~km, which correspond to one modification of
APR EOS \citep*{akmaletal98}, a typical neutron star model
suitable for analysing the observations of the Cas~A neutron star.

If the temperature decline is due to cooling, it is inevitably
small (over a limited observation history) so that in reality one
obtains (detects) some mean value of $T_{\rm s}$ and the value of
$s=s_{\rm d}$ (equation \ref{e:defs}).  To perform a full
analysis, we need models of heat-blanketing envelope. We will
employ the same carbon--iron envelope models as were used by
\citet{yakovlevetal11}, with $\Delta M/{\rm M_\odot}=0$,
$10^{-11}$ and $10^{-8}$ mass of carbon. The nonredshifted
temperature at the bottom of heat-blanketing envelope in our
example is $T_{\rm b}=3.58,\,2.59$ and $2.05\times 10^8$~K for the
three selected amounts of carbon and the redshifted temperature of
the isothermal interior is $T_{\rm d}=2.74,\, 1.98$ and
$1.57\times 10^8$~K, respectively. The envelope with carbon is
more transparent to heat, making the star with this envelope
colder inside than the star of the same surface temperature but
with the iron envelope.

The basic parameters of the
neutron star model with $\Delta M=0$ are collected in Table \ref{tab:model}.

\begin{table}
\caption[]{An example of Cas A neutron star model (iron heat
blanket): the employed stellar mass $M$, radius $R$, age $t_{\rm
d}$, surface temperature $T_{\rm s}$, redshifted internal
temperature $T_{\rm d}$ (for the iron envelope) and the standard
candle neutrino cooling rate $\ell_{\rm SC d}$ (for these $M$, $R$
and $T_{\rm d}$).}
\begin{tabular}{lc@{\hspace{1ex}}cc@{\hspace{1ex}}cc}
\hline
~$M$~  & ~$R$~  & ~$ t_{\rm d}$~  & ~$T_\mathrm{s}$~  &
$T_{\rm d}$ & $\ell_{\rm SCd}$ \\
~${\rm M_\odot}$~  & ~km~  & ~yr~  & ~MK~ & ~MK~  & ~Myr/K~  \\
 \hline\hline
1.65  & 11.8   &  330   & 2.0  & 274   &  0.138      \\
\hline
\end{tabular}
\label{tab:model}
\end{table}

\subsection{Analysing data from measured values of~$s_{\rm d}$}
\label{s:selfsim}

First, we describe which information on neutron star physics can
be extracted from the detected $s_{\rm d}$. Since $s_{\rm d}$ is
almost independent of the model of the heat-blanketing envelope,
all results of this analysis also possess this property.

Let us take the theoretical expression (\ref{e:s-calc}) for
$s(\tau,\delta)$, equate it to the detected $s_{\rm d}$, and
consider it as an equation to be solved. For any $\tau_{\rm
d}=T_{\rm d}/T_{\rm C}<1$, we can easily solve it and find the
value of $\delta$, which gives us a cooling solution for chosen
$s_{\rm d}$ and $\tau_{\rm d}$. In this way, we construct a family
(continuum) of solutions parametrized by the values of $\tau_{\rm
d}$.

\begin{figure}
\includegraphics[width=0.45\textwidth]{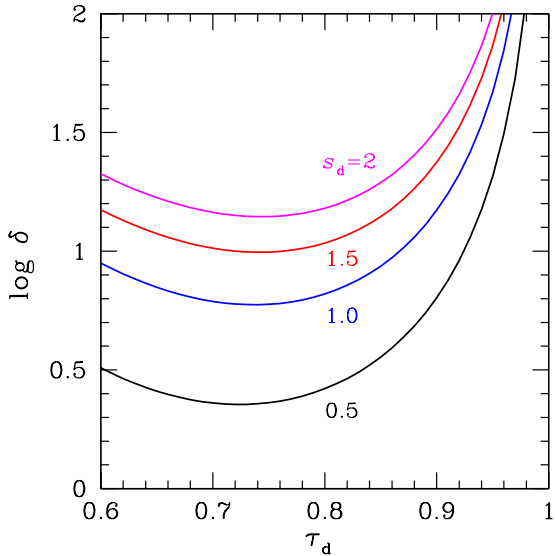}
\caption{(Color on line) Four families of cooling solutions which
give predetermined values $s_{\rm d}=$0.5, 1, 1.5 and 2 in the
present epoch $t=t_{\rm d}$. Any solution is parametrized by
$\tau_{\rm d}=T_{\rm d}/T_{\rm C}$. The figure shows the parameter
$\delta$ of neutron superfluidity strength for any solution.
\label{fig:appdelvsT} }
\end{figure}

\begin{figure}
\includegraphics[width=0.48\textwidth]{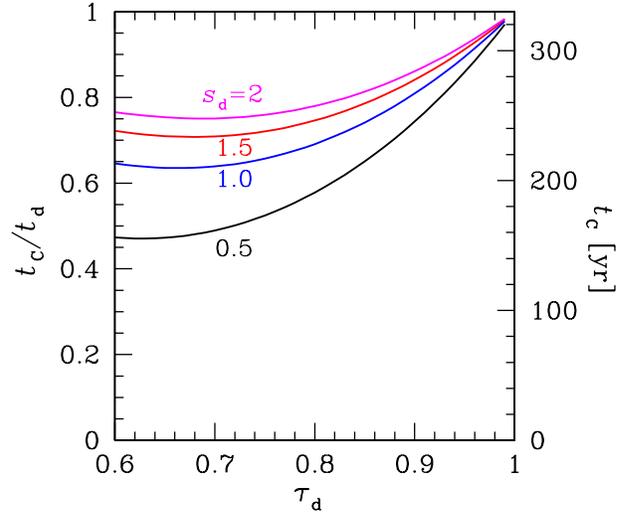}
\caption{(Color on line) The same four families of cooling
solutions for $s_{\rm d}=$2, 1.5, 1 and 0.5 as in Fig.\
\ref{fig:appdelvsT}. The figure presents $t_{\rm C}/t_{\rm d}$
(left vertical scale) for any $\tau_{\rm d}=T_{\rm d}/T_{\rm C}$.
For illustration, the right vertical scale gives $t_{\rm C}$ for
the Cas A neutron star.\label{fig:apptvsT} }
\end{figure}

The results are illustrated in Figs.\ \ref{fig:appdelvsT} and
\ref{fig:apptvsT}. In these figures, we display four families of
such solutions which correspond to $s_{\rm d}=$0.5, 1, 1.5 and 2.
Figs.\ \ref{fig:appdelvsT} and  Fig.\ \ref{fig:apptvsT} show,
respectively, the values of $\delta$ and $t_{\rm C}/t_{\rm d}$
versus $\tau_{\rm d}$. In addition, the right vertical scale  in
Fig.\ \ref{fig:apptvsT} presents the time $t_{\rm C}$ of neutron
superfluidity onset in the Cas A neutron star. For higher $s_{\rm
d}$, one naturally needs stronger neutrino cooling due to Cooper
pairing neutrino emission (larger $\delta$). For a given $s_{\rm
d}$, the lowest $\delta$ corresponds to the vicinity of $\tau_{\rm
d}=T_{\rm d}/T_{\rm C} \approx 0.77$ (to the peak of Cooper
pairing neutrino outburst in the present epoch). The limit of
$\tau_{\rm d} \to 1$ is equivalent to $T_{\rm C} \to T_{\rm d}$
[the epoch $t_{\rm C}$ of neutron superfluidity onset approaches
the present (detection) epoch $t_{\rm d}$].

Let us emphasize that families of cooling solutions for fixed
$s_{\rm d}$ are really {\it selfsimilar and universal}. They are
not only {\it independent of the model for the heat-blanketing
envelope} but {\it independent also of the neutron star model
(mass, radius, the EOS) as well as of absolute value of the
surface
 temperature} $T_{\rm s}$. {All these dependences are encapsulated
in the values of introduced dimensionless parameters. If, by any
chance, one of them is known} (for instance, $\delta$, from a
given model of neutron superfluidity), then one can use this known
value and find $\tau_{\rm d}$. In this case, one would select a
unique solution of the cooling problem (or a pair of them) from
the entire family. Otherwise, one should face the family of
solutions with different $\tau_{\rm d}$.

\subsection{Solutions with particular $\tau_{\rm d}$}

Here, we describe which additional information can be inferred
from observations provided {particular} $\tau_{\rm d}$ is
{selected}. Let us take the corresponding cooling solution from
the family described in Section \ref{s:selfsim} (with the iron
heat-blanketing envelope as an example). Using the values of
$\tau_{\rm d}$ and $T_{\rm d}$, we immediately find the maximum
critical temperature of neutron superfluidity, $T_{\rm C}$. Using
equation (\ref{e:cool-SF1}) as well as specific values of
$\tau_{\rm d}$ and $\delta$, one can determine $t_{\rm d}/t_{\rm
C}$ and obtain the time $t_{\rm C}$ of neutron superfluidity
onset.

 According to \citet{yakovlevetal11}, it is convenient to describe
 the  neutrino emission level via the ratio
\begin{equation}
f_\ell=\ell(T)/\ell_{\rm SC}(T)
\label{e:fell}
\end{equation}
of the neutrino cooling rate $\ell(T)$ of our star to the neutrino
cooling rate $\ell_{\rm SC}(T)$ of the standard candle  (the star
of the same $M$ and $R$ which cools slowly via modified Urca
process) at the same internal temperature. At $t<t_{\rm C}$, the
cooling is slow and the factor $f_\ell=f_{\ell 0}$ is just a
number (independent of $T$) which reflects the neutrino cooling
level prior to superfluidity onset. This level can be determined
from $t_{\rm C}$ and $T_{\rm C}$ as
\begin{equation}
    f_{\ell0}=[T_{\rm SC}(t_{\rm C})/T_{\rm C}]^6,
\label{e:f-SC}
\end{equation}
where $T_{\rm SC}(t)$ is given by equation 14 of
\citet{yakovlevetal11}.

Now, one has everything at hand to fully reconstruct the cooling
history of the star in absolute and dimensionless variables for
any cooling solution of the family. Any solution is characterized
by the parameters $f_{\ell 0}$, $\delta$ and $T_{\rm C}$ which
determine the efficiency of neutrino cooling. The value of
$f_{\ell 0}$ contains all the information on the neutrino cooling
in the early epoch when neutron superfluidity in the core is
absent. The values of $T_{\rm C}$ and $\delta$ describe neutron
superfluidity and the neutrino cooling rate after the neutron
superfluidity onset. This analysis is independent of a specific
model of neutron star (with nucleon core). Specific physical
models which agree with the inferred results can be analysed at a
later stage.

\begin{figure}
\includegraphics[width=0.45\textwidth]{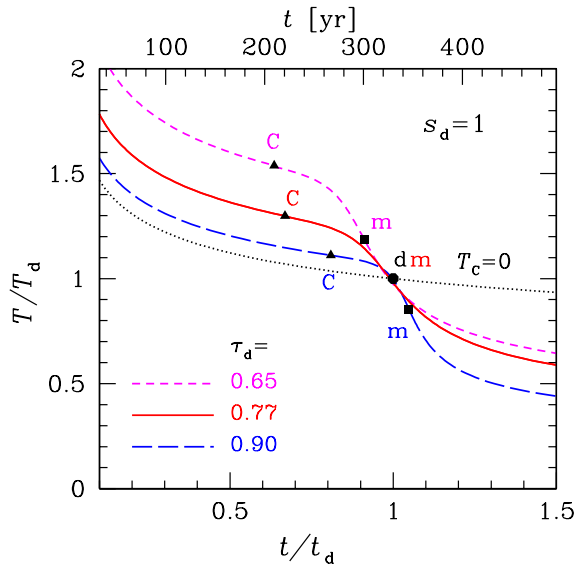}
\caption{(Color on line) Evolution of internal temperature
$T/T_{\rm d}$ versus $t/t_{\rm d}$ for  in the 1.65$\,{\rm
M_\odot}$ neutron star in Cas~A with iron heat-blanketing
envelope, nonredshifted surface temperature $T_{\rm sd}=2$ MK
($T_{\rm d}=274$~MK) and $s_{\rm d}=1$  at $t=t_{\rm d}=330$ yr
for the three values $\tau_{\rm d}=T_{\rm d}/T_{\rm C}$=0.65, 0.77
and 0.9 (short-dashed, solid and long-dashed lines, respectively).
Moments of time `C' when neutron superfluidity sets in are denoted
by triangles. Moments `m' of maximum Cooper pairing neutrino
emission rate ($\tau=0.77$) are labelled by squares, while `d'
refers to the moment of observation. The dotted line shows the
evolution of the star without neutron superfluidity. The upper
horizontal scale gives real time $t$. See text for details.
\label{fig:coolsim} }
\end{figure}

\begin{figure}
\includegraphics[width=0.45\textwidth]{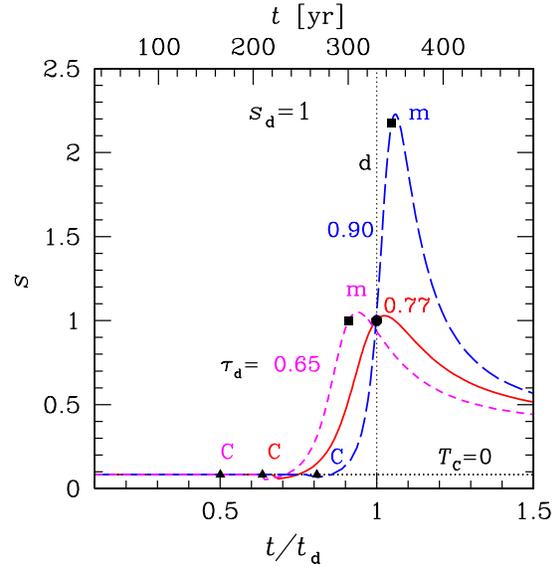}
\caption{(Color on line) Evolution of surface temperature decline
$s$, equation (\ref{e:s-calc}), of the Cas A neutron star
calculated for the same three cooling scenarios ($\tau_{\rm
d}$=0.65, 0.77 and 0.9) as in Fig.\ \ref{fig:coolsim}. The
vertical dotted line shows the present-time epoch. See text for
details. \label{fig:ssim} }
\end{figure}

\begin{figure}
\includegraphics[width=0.45\textwidth]{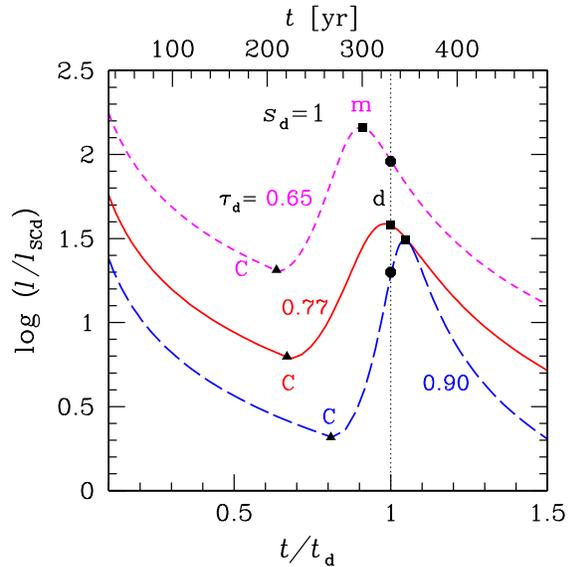}
\caption{(Color on line) Evolution of neutrino cooling rate $\ell$
(in units of $\ell_{\rm SCd}$ of standard neutrino candle at $t=t_{\rm d}$,
Table \ref{tab:solutions})
of the Cas A neutron star for the same three
cooling scenarios as in Figs.\ \ref{fig:coolsim} and
\ref{fig:ssim}. The vertical dotted line
shows the present-time epoch.
See text for details. \label{fig:lsim} }
\end{figure}

\begin{figure}
\includegraphics[width=0.45\textwidth]{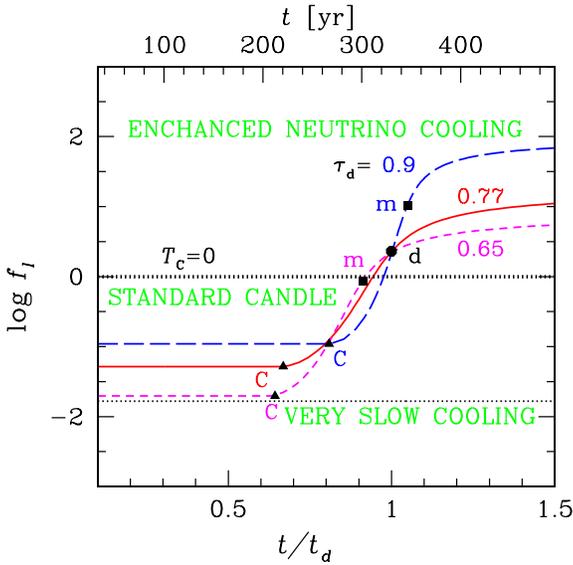}
\caption{(Color on line) Evolution of neutrino cooling rate
$f_\ell$ (in standard candles) of the Cas A neutron star for the
same three cooling scenarios ($\tau_{\rm d}$=0.65, 0.77 and 0.9)
as in Figs.\ \ref{fig:coolsim}--\ref{fig:lsim}. {The horizontal
thicker dotted line $f_\ell=1$ refers to the standard neutrino
candle; the thinner dotted line $f_\ell=1/60$ is the estimated
lowest theoretical rate.} See text for details. \label{fig:flsim}
}
\end{figure}

\begin{figure}
\includegraphics[width=0.45\textwidth]{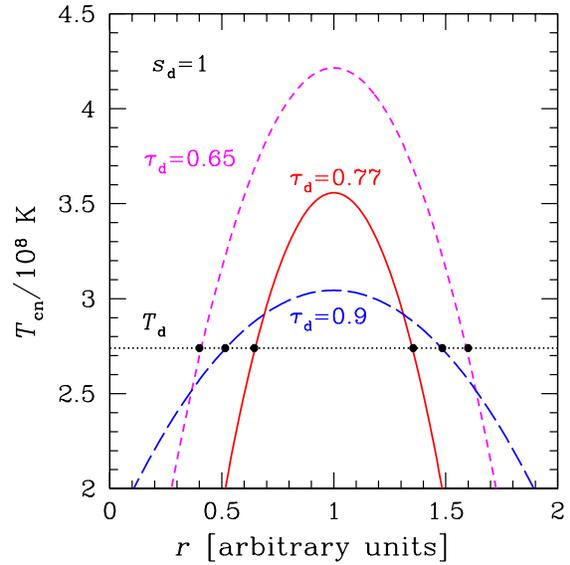}
\caption{(Color on line) Examples of the three neutron
superfluidity profiles $T_{\rm cn}(r)$ over the neutron star core
which produce the cooling solutions for $\tau_{\rm d}$=0.65
(short-dashed line), 0.77 (solid line) and 0.90 (long dashes),
respectively. The radial coordinate $r$ is in arbitrary units. The
dotted horizontal line is the core temperature in the present
epoch. Filled dots show the boundaries of superfluid layers in the
three cases. \label{fig:tcr} }
\end{figure}

\begin{figure}
\includegraphics[width=0.45\textwidth]{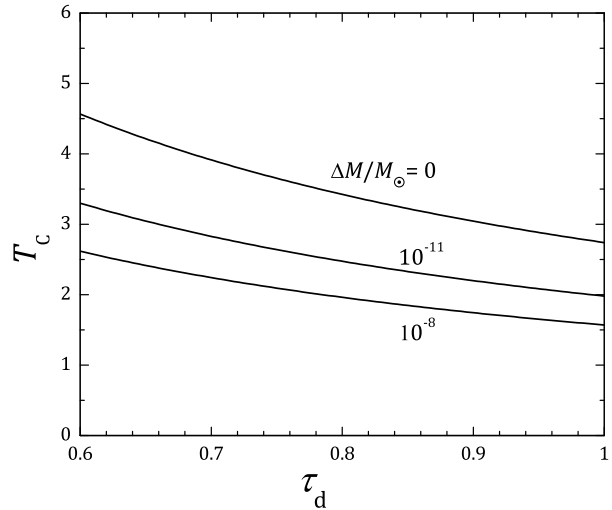}
\caption{ Maximum (redshifted) critical temperature  for neutron
superfluidity $T_{\rm C}$ (which is independent of $s_{\rm d}$)
versus $\tau_{\rm d}$ in the 1.65\,${\rm M_\odot}$ model of the
Cas A neutron star at three values of $\Delta M/{\rm M_\odot}$
shown near the lines ($\Delta M$ being the carbon mass in the
heat-blanketing envelope; see text for details). \label{fig:appTc}
}
\end{figure}

\begin{figure}
\includegraphics[width=0.45\textwidth]{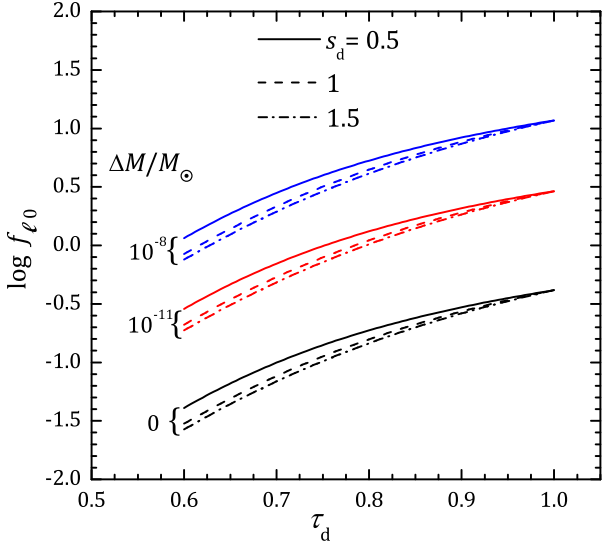}
\caption{(Color on line) Neutrino cooling rate $f_{\ell 0}$ (in
standard candles) for the Cas A neutron star model before the
neutron superfluidity onset for three families of cooling
solutions ($s_{\rm d}=$0.5, 1, 1.5) versus $\tau_{\rm d}$ at the
three values of $\Delta M/{\rm M_\odot}$ (the same as in Fig.\
\ref{fig:appTc}) indicated in the plot. \label{fig:appafl} }
\end{figure}

For illustration, consider three cooling solutions for $s_{\rm d}=1$
at $\tau_{\rm d}=0.65$, 0.77 and 0.9. Some parameters of these
solutions are listed in Table \ref{tab:solutions}, where $\Delta
t_{\rm C}=t_{\rm d}-t_{\rm C}$ and $\Delta t_{\rm m}=t_{\rm
d}-t_{\rm m}$  (where $t_{\rm m}$ is the time corresponding to
$\tau=\tau_{\rm m}=0.77$).

\begin{table}
\caption[]{Three cooling solutions for the Cas A neutron star model
with iron heat blanket and $s_{\rm d}=1$. See text for details.}
\begin{tabular}{c c c c c c}
\hline
$\tau_{\rm d}$  & $\delta$ & $\Delta t_{\rm C}~^{(a)}$ & $T_{\rm C}$
& $\Delta t_{\rm m}~^{(b)}$  & $s_{\rm max}$ \\
  &      &  yr   & MK   & yr &  \\
 \hline\hline
0.65  & 7.046   & 120  &  422  & 29   &  1.05  \\
0.77  & 6.111     & 109 & 356  &  0  &  1.00  \\
0.90  & 14.88     & 63  & 304  & --16 & 2.23 \\
\hline
\end{tabular}\\
$^{(a)}\;\Delta t_{\rm C}=t_{\rm d}-t_{\rm C}$\\
$^{(b)}\;\Delta t_{\rm m}=t_{\rm d}-t_{\rm m}$
\label{tab:solutions}
\end{table}

Fig.\ \ref{fig:coolsim} shows the internal thermal evolution
($T/T_{\rm d}$ versus $t/t_{\rm d}$ or $t$, lower or upper
horizontal scales, respectively) of the Cas A neutron star
assuming the iron heat-blanketing envelope ($T_{\rm d}= 2.74
\times 10^8$ K). It is additionally assumed that at $t=t_{\rm d}$
the surface temperature decline is $s_{\rm d}=1$. The surface
temperature behaves approximately as $T_{\rm s}(t)\approx T_{\rm
s}(t_{\rm d})(T(t)/T_{\rm d})^{0.5}$. We show the three cooling
curves parametrized by $\tau_{\rm d}$=0.65, 0.77 and 0.9 (the
short-dashed, solid and long-dashed curves, respectively). These
are three possible cooling scenarios (among continuum of others),
which give the same temperature and temperature decline of the
star in the observation epoch (filled point `d'). They differ by
$\tau_{\rm d}$, that is by the maximum critical temperature
$T_{\rm C}$ for neutron superfluidity in the core (Table
\ref{tab:solutions}). The moments of time `C', when neutron
superfluidty appears in the core (120, 109 and 63 yr ago, for
$\tau_{\rm d}=0.65$, 77 and 0.9, respectively), are marked by
triangles on the cooling curves. The higher $T_{\rm C}$, the
earlier it should appear. By squares (points `m'), we mark the
moments of maximum Cooper pairing neutrino emission rate
($\tau=0.77$, approximately, the maximum of neutrino outburst). In
case $\tau_{\rm d}=0.65$, this maximum is reached 29 yr before the
observation epoch, while in case $\tau_{\rm d}=0.9$, it occurs 16
yr after the observation epoch. If  $\tau_{\rm d}=0.77$, it occurs
just now. By the dotted line, we plot the cooling curve calculated
neglecting neutron superfluidity and slightly adjusting the
modified Urca neutrino emission level to give  the same current
temperature $T_{\rm d}$ of the star,  as other curves.  This
cooling gives too low $s=1/12$ and would be not detectable in real
time for the Cas A neutron star.

Fig.\ \ref{fig:ssim} presents the evolution of the surface
temperature decline $s$ for the same three cooling scenarios of
the Cas A neutron star as in Fig.~\ref{fig:coolsim}, and for the
standard neutrino candle without neutron superfluidity. The
notations are the same as in Fig.~\ref{fig:coolsim}. In the
scenario with $\tau_{\rm d}=0.65$, the maximum  $s_{\rm
max}\approx 1.05$ is reached prior to the detection epoch (about
20 yr ago), so that $s(t)$ decreases with time during the
detection epoch. This scenario is qualitatively consistent with
those suggested by \citet{pageetal2011} and
\citet{shterninetal2011}. In the scenario with $\tau_{\rm d}=0.9$,
$s(t)$ has not yet reached its maximum at the present epoch.
Accordingly, $s(t)$ sharply increases with $t$ and will reach
maximum $s_{\rm max}\approx 2.23$ in about 20 yr from now. At
$\tau_{\rm d}=0.77$, the maximum value $s_{\rm max}\approx 1$ is
reached just now. Therefore, $s(t)$ should decrease in time but in
the next 20 years the decrease should be very slow. Let us remark
that the maxima of $s(t)$ are close to but do not coincide with
the maxima of $\ell_{\rm CP}(t)$. The stronger the Cooper pairing
neutrino emission, the better the coincidence.

Fig.\  \ref{fig:lsim} displays the evolution of the neutrino
cooling rate $\ell$ (in the units of the rate $\ell_{\rm
SCd}=\ell_{\rm SC}(t_{\rm d})=0.138$ MK yr$^{-1}$ for the standard
candle in the present epoch, see Table~\ref{tab:model}) for the
same three scenarios as in Figs.\ \ref{fig:coolsim} and
\ref{fig:ssim}. Prior to the neutron superfluidity onset, we have
$\ell(T) \propto T^7$. With increasing $\tau_{\rm d}$, we need
lower $\ell(t)$. After superfluidity onset, the rate is enhanced
by the neutrino outburst with the maximum before the detection
epoch (at $\tau_{\rm d}=0.65$), just now ($\tau_{\rm d} =1$) or
afterwards ($\tau_{\rm d}=0.9$). Again, the maxima of $\ell(t)$ do
not exactly coincide with the maxima of $\ell_{\rm CP}(t)$ (with
points `m') but the coincidence becomes better for stronger Cooper
pairing neutrino emission.

Fig.\ \ref{fig:flsim} demonstrates the evolution of  neutrino
cooling rate $f_\ell(t)$, expressed in standard neutrino candles
according to equation~(\ref{e:fell}), for the same three cooling
scenarios ($\tau_{\rm d}$=0.65, 0.77 and 0.9) as in Figs.\
\ref{fig:coolsim}--\ref{fig:lsim}.  Note that $f_\ell=1$ ($\log
f_\ell=0$) refers to the standard neutrino candle, $\log f_\ell
\gtrsim 2$ to rather enhanced neutrino cooling and $\log f_\ell
\lesssim -2$ to unrealistically slow cooling. We see that prior to
the neutron superfluidity onset (before triangles),
$f_\ell=f_{\ell 0}$ is constant and rather low. This agrees with
numerical simulations of the Cas A neutron star cooling by
\citet{pageetal2011} and \citet{shterninetal2011}. With increasing
$\tau_{\rm d}$ (or, equivalently,  decreasing $T_{\rm C}$), one
needs higher $f_{\ell 0}$. At very low $\tau_{\rm d}$, one would
need too slow neutrino cooling rate before the neutron
superfluidity onset. {For instance, $f_{\ell 0}$ cannot be
$\lesssim 1/60$ (thinner horizontal dotted line in Fig.\
\ref{fig:flsim}) because proton superfluidity cannot produce too
strong reduction of the slow cooling rate with respect to the
modified Urca rate (there are always processes, such as
neutron--neutron or electron--electron neutrino bremsstrahlung,
which do not involve protons; they are almost insensitive to the
presence of proton superfluidity). Such cases are unrealistic and
should be disregarded.}

After the neutron superfluidity is switched on, $f_\ell(t)$ in
Fig.\ \ref{fig:flsim} grows up (during the neutrino outburst) and
then has the tendency to saturate at much higher level than at the
initial cooling stage. This saturation reflects the fact that the
neutrino emission due to Cooper pairing of neutrons in a developed
neutron superfluidity has the same temperature dependence as
the standard candle ($\ell
\propto \tau^7$) but can be substantionally enhanced with respect to the
standard candle (\citealt{pageetal04}, \citealt{gusakovetal04}).
The higher $\tau_{\rm d}$, the larger enhancement. Very large enhancements
$f_\ell\gtrsim 10^2$ are unrealistic and should be disregarded.

Fig.\ \ref{fig:tcr} demonstrates possible $T_{\rm cn}(r)$ profiles
as a function of radial coordinate $r$ within the neutron star
core. These profiles can realize cooling solutions with $\tau_{\rm
d}=0.65$, 0.77 and 0.90 (short-dashed, solid and long-dashed
lines, respectively). The profiles are approximated by inverted
parabolas whose maxima are at the same position $r_\mathrm{C}$ in
the star. The maximum heights  $T_\mathrm{C}$ and the parameters
$\delta$ have already been determined (Table \ref{tab:solutions}).
The characteristic widths $\Delta r_\mathrm{C}$ of the parabolas
are found from equation (\ref{e:deltasim}) [up to a joint
normalization factor determined by the coefficient $A(r_{\rm C})$;
to avoid cumbersome calculation of $A(r_{\rm C})$, we plot radial
coordinates in arbitrary units]. The horizontal dotted line shows
the present-day temperature $T_\mathrm{d}$ in the core. Neutron
superfluidity exists at those $r$ at which $T_{\rm cn}(r)\geq
T_{\rm d}$. The boundaries of superfluid layers are marked by
dots. For the solution with $\tau_{\rm d}=0.65$, the $T_{\rm
cn}(r)$ profile is sufficiently high. For $\tau_{\rm d}=0.77$, it
is smaller, while for $\tau_{\rm d}$, it is even smaller but
wider. The increased width is needed to obtain large
$\delta=14.88$. Naturally, Fig.\ \ref{fig:tcr} presents only some
examples of $T_{\rm cn}(r)$ profiles. The same cooling solutions
can be realized with other profiles [different values of $r_{\rm
C}$ and $A(r_{\rm C})$ in equation (\ref{e:deltasim})] which
result in the same $\delta$.

\subsection{Effects of carbon heat-blanketing envelope and
different~$s_{\rm d}$}

So far, we have analysed cooling models of the Cas A neutron star
only at $s_{\rm d}=1$ and with standard heat-blanketing envelopes
made of iron. Let us outline the effects of possible carbon heat
blankets and different present-day slopes $s_{\rm d}$ of the
cooling curves.

Fig.\ \ref{fig:appTc} presents the maximum critical temperature
$T_{\rm C}$ for neutron superfluidity in the core of the
1.65\,${\rm M_\odot}$ neutron star (the same as considered
throughout this paper) for different cooling solutions
parametrized by $\tau_{\rm d}$. They are apparently determined by
$T_{\rm d}$ being independent of $s_{\rm d}$. The upper line
corresponds to the iron heat blanket while two lower lines refer
to the heat blankets containing $\Delta M$ = $10^{-11}$ and
$10^{-8}\,{\rm M_\odot}$ of carbon, respectively. The presence of
carbon makes the heat-blanketing envelope more heat transparent
and reduces the critical temperature $T_{\rm C}$ required to
satisfy cooling solutions. The  reduction is seen to be quite
substantial.

Fig.\ \ref{fig:appafl} shows logarithm of the neutrino cooling
rate $f_{\ell 0}$ (in standard candles) prior to the onset of
neutron superfluidity for the cooling solutions as a function of
$\tau_{\rm d}$. We show three groups of curves, again for $\Delta
M/{\rm M_\odot}$=0, $10^{-11}$ and $10^{-8}$ (from top to bottom).
For each amount of carbon $\Delta M$, we present the solutions for
$s_{\rm d}=0.5$ (solid lines), 1 (dashed lines) and 1.5
(dot-dashed lines). {Higher amount of carbon leads to lower
$T_{\rm d}$ and} the solutions require larger $f_{\ell 0}$; for
larger $s_{\rm d}$, they require lower $f_{\ell 0}$. It is seen
that $f_{\ell 0}$ is rather insensitive to $s_{\rm d}$ in the
given $s_{\rm d}$ interval.

One can see that for the cases of iron envelope and envelope with
$10^{-11}\;{\rm M_\odot}$ of carbon, the slow cooling rate prior
to the neutron superfluidity onset should be lower than for the
standard candle. This lowering can be provided by strong proton
superfluidity in the neutron star core. With the growth of
$\tau_{\rm d}$, the {required} lowering is smaller. Taking highest
amount of carbon ($\Delta M = 10^{-8}\;{\rm M_\odot}$), one will
need the standard cooling rate, or even enhanced cooling at
$t<t_{\rm C}$.

\begin{figure}
\includegraphics[width=0.45\textwidth]{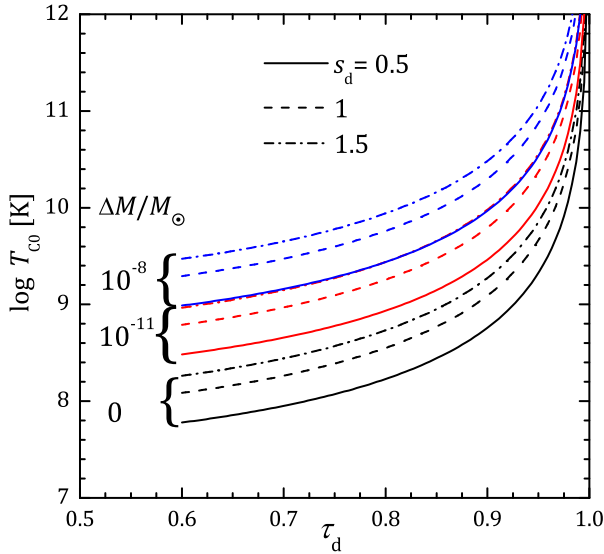}
\caption{(Color on line) Parameter $T_{\rm C 0}$ versus $\tau_{\rm
d}$ for three families of cooling solutions ($s_{\rm d}=$0.5, 1,
1.5) and  three values of $\Delta M/M_\odot$ (as in
Fig.~\ref{fig:appafl}). \label{fig:TC0} }
\end{figure}

These solutions can be constrained further by taking into account
natural physical restrictions. {Recall that $f_{\ell 0}$ should be
$\gtrsim 1/60$.} For the iron envelope model, this invalidates
cooling solutions with rather small $\tau_{\rm d}$ and large
$s_{\rm d}$. On the other hand, $f_{\ell 0}$ cannot be arbitrarily
large as this would require unphysically strong neutrino outburst
after superfluidity onset to reach the same $s_{\rm d}$ (the same
$\delta$) at the present epoch.

According to equation~(\ref{e:def-delta}), $\delta$ is inversely
proportional to the value $\ell_{\rm C}$ which determines the
neutrino emission rate prior to neutron superfluidity onset. In
addition, as seen from (\ref{e:deltasim}), $\delta\propto T_{\rm
C}^{-1}$. Therefore, it is instructive to introduce the new
parameter
\begin{equation}
  T_{\rm C0 }= f_{\ell 0} T_{\rm C}\, \delta
\label{T_C0}
\end{equation}
instead of $\delta$. Its dimension is temperature but it
characterizes the volume of the region occupied by neutron
superfluidity in the star. {Because of the factor $f_{\ell 0}$ in
equation~(17)  the dependence of $T_{\rm C0}$ on the neutrino
emission level prior the neutron superfluidity onset is
eliminated.}
 If we fix $r_{\rm C}$ but
increase $\delta r_{\rm C}$, we would amplify $T_{\rm C0} \propto
\delta r_{\rm C}$ (see equation~(\ref{e:deltasim})). In
Fig.~\ref{fig:TC0}, we plot $T_{\rm C0}$ versus $\tau_{\rm d}$ for
the selected values of $s_{\rm d}$ and $\Delta M$. This figure
demonstrates the dependence of the peak's width $\delta r_{\rm C}$
on its height $T_{\rm C}$ for selected families of cooling
solutions characterized by $s_{\rm d}$ and $\tau_{\rm d}$. For a
given $s_{\rm d}$, higher $\tau_{\rm d}$ would require wider
$T_{\rm cn}(r)$ peaks (larger $\delta r_{\rm C}$). Equally, at a
fixed $\tau_{\rm d}$, higher $s_{\rm d}$ require larger $\delta
r_{\rm C}$. Naturally, $\delta r_{\rm C}$ is limited by the size
of the neutron star core so that $T_{\rm C0}$ cannot be
arbitrarily large.

According to cooling simulations, realistic models of neutron
superfluidity correspond to $T_{\rm C0}\lesssim 3\times 10^{8}$~K.
As an example, consider the Cas~A neutron star at $s_{\rm d}=1$.
From Fig.~\ref{fig:TC0}, we find $\tau_{\rm d}\lesssim 0.75$ for
the iron heat blanket. Then, Fig.~\ref{fig:appafl} implies
 $f_{\ell 0}\lesssim 0.1$. Lower values of $\tau_{\rm d}$ require
lower $T_{\rm C0}$, i.e. smaller volume occupied by neutron
superfluidity or weaker efficiency of the Cooper paring neutrino
emission. However, the price to pay is that the $f_{\ell 0}$
should also be lower. Taking into account that $f_{\ell 0}$ cannot
be too small and
 $\tau_{\rm d}\gtrsim 0.6$, we obtain an approximate constraint $3.5\times
10^8 \lesssim T_{\rm C} \lesssim 4.5\times 10^8$~K. This in turn
means that real (non-redshifted) maximum critical temperature of
neutrons lies in the range $\sim (5-8)\times 10^8$~K depending on
the position $r_{\rm C}$ of the maximum critical temperature
$T_{\rm cn}$ inside the core for our 1.65~${\rm M_\odot}$ neutron
star model. The restriction of low $T_{\rm C0}$ forbids
significant amount of carbon in the envelope in our example.
According to Figs.~\ref{fig:appafl} and \ref{fig:TC0}, the values
$\Delta M\gtrsim 10^{-11}\,{\rm M_\odot}$ are inconsistent with
the observations if $s_{\rm d}=1$. If $s_{\rm d}=0.5$, then
$\Delta M\sim 10^{-11}\,{\rm M_\odot}$ is allowed but with fine
tuning of the parameters to obey $\tau_{\rm d}\lesssim 0.6$. For
this solution we need $f_{\ell 0}<0.3$  (Fig.\ \ref{fig:appafl}),
i.e. we also need (not very strong) proton superfluidity in the
core. Note that numerical cooling solutions with $\tau_{\rm
d}<0.6$ follow the general trend of
Figs.~\ref{fig:appafl}--\ref{fig:TC0}. However, strictly speaking,
our simple analytical formalism is inapplicable at such low
$\tau_{\rm d}$. On the other hand, in case $s_{\rm d}=1$ neutron
superfluidity should be inevitably rather strong, so that $T_{\rm
C0}>10^{8}$~K.

Note that the limit  $T_{\rm C0} \lesssim 3\times 10^8$~K was
estimated using the same model for treating the collective effects
on the efficiency of Cooper paring neutrino emission as adopted by
\citet{pageetal09, pageetal2011} and \citet{shterninetal2011}
(their reduction factor $q=0.76$ of the neutrino emissivity by the
collective effects). In this model, the emission in the vector
channel is fully suppressed by the collective effects, while the
emission in the axial vector channel remains unchanged. According
to \citet{leinson10}, collective effects may actually lower the
neutrino emission efficiency four times more ($q=0.19$ in
\citealt{shterninetal2011}). The latter case corresponds to the
restriction $T_{\rm C0}\lesssim 0.75 \times 10^8$~K. According to
Fig.~\ref{fig:appafl}, it is impossible to get $s_{\rm d}=1$ with
such $T_{\rm C0}$. If, however, the cooling of the Cas~A neutron
star is slower, with $s_{\rm d}\sim 0.5$, then we again can
explain the observations, provided $\tau_{\rm d}<0.66$, even for
such a low efficiency of Cooper pairing neutrino emission. In this
case, we need strong proton superfluidity (low $f_{\ell 0}$), and
a small amount of carbon in the heat-blanketing envelope.

Of course, the described procedure of data analysis is idealized.
All observables ($M$, $R$, $T_{\rm s}$, $t_{\rm d}$) are always
determined with some uncertainties.
This biases the
analysis of the data and introduces uncertainties into final
results.

We have compared some analytic cooling solutions with those
obtained with our cooling code \citep{gnedinetal01} and found out
impressive agreement. Notice, however, that, according to the
exact solutions, the appearance of neutron superfluidity and the
associated neutrino outburst slightly violate isothermality of the
core (e.g., \citealt{shterninetal2011}) but this violation has no
noticeable effect on the cooling curves.

\section{Conclusions}
\label{s:conclusions}

We have analysed the cooling of a neutron star with the thermally
relaxed nucleon core at the neutrino cooling stage ($10^2 \lesssim
t \lesssim 10^5$ yr). For simplicity, we have considered neutron
star models where direct Urca process does not operate. We have
assumed further that the star has strong proton superfluidity in
the core, which appears at the early cooling stage, and moderately
strong (triplet-state) neutron superfluidity which appears later,
when the internal temperature of the star, $T$, falls below
$T_{\rm C}$, the maximum critical temperature for neutron
superfluidity over the stellar core (equation (\ref{e:TC})).
Therefore, the star cools slowly before the neutron superfluidity
onset ($T>T_{\rm C}$, $t<t_{\rm C}$) but its cooling is
accelerated later by the appearance of neutron superfluidity and
associated outburst of the neutrino emission due to Cooper pairing
of neutrons.

Our analysis is based on the results by \citet{gusakovetal04}
according to which at $T$ not much lower than $T_{\rm C}$, the
neutrino luminosity $L_\nu^{\rm CP}(T)$ due to Cooper pairing of
neutrons has a universal form. We show that at these temperatures,
$0.6\,T_{\rm C} \lesssim T <T_{\rm C}$, the neutrino cooling rate
$\ell(T)=L_\nu(T)/C(T)$ is approximated by the simple expression
(\ref{e:total-ell}), and the cooling problem is solved in a closed
analytic self-similar forms (\ref{e:slowcool}) and
(\ref{e:cool-SF}). Any solution can be parametrized by the values
of $\tau_{\rm d}=T_{\rm d}/T_{\rm C}$ and $s_{\rm d}$ (the slope
of the cooling curve in the present epoch, $t=t_{\rm d}$).
Formally, for a fixed $s_{\rm d}$, there  exists a continuum of
solutions which differ by the values of $\tau_{\rm d}$. We have
analysed the properties of these solutions and the methods to
select physically sound ones. In the essence, the solutions differ
by the profiles $T_{\rm cn}(\rho)$ and $T_{\rm cp}(\rho)$ of
critical temperatures for neutron and proton superfluidity in the
neutron star core. However, our analytic approach allows one to
describe the effects of these superfluidities on the neutron star
cooling by two dimensionless parameters, $f_{\ell 0}$ [equation
(\ref{e:f-SC}), reflects the neutrino cooling rate prior to
neutron superfluidity onset, regulated by proton superfluidity],
and $\delta$  [equation (\ref{e:def-delta}), characterizes the
efficiency of neutrino outburst due to neutron superfluidity]. We
have described how to infer allowable values of these parameters
from observations of neutron stars whose cooling in real time is
observed, using the Cas A neutron star as the only example known
today.

The advantage of our method is that it gives all possible
solutions of the cooling problem. One can analyse them and
determine all the values of the parameters (particularly, $f_{\ell
0}$ and $\delta$); whereas physical models of superfluidity
[$T_{\rm cn}(\rho)$ and $T_{\rm cp}(\rho)$] can be investigated at
the later stage. In this way, we have extended the model
independent method of data analysis of cooling neutron stars
suggested by \citet{yakovlevetal11}. {The latter authors developed
this method for slowly cooling neutron stars.} We have included a
more complicated case of neutron superfluidity onset.

The analytic solution can be used to interpret observations of
cooling neutron stars in real time (when one can measure the
surface temperature of the star $T_{\rm s}$ and the rate $s_{\rm
d}$ of its decline). We have described the procedure (Section
\ref{s:analysis}) how to interpret such observations, to
reconstruct the cooling history of the neutron star and predict
its future cooling behaviour (for future observational tests). We
have presented {examples} of such interpretations for the Cas~A
neutron star. {In particular, one needs to suppress the neutrino
emission  prior to the neutron superfluidity onset below the
modified Urca level even if the rapid cooling in real time at the
present epoch is twice slower than estimated by
\citet{heinkeho10}. Moreover, we have shown that large amount of
carbon in the heat-blanketing envelope is inconsistent with
observations of this object. }

Because the observations of this star are still a subject of
debates (Section \ref{s:CasA}), one should be ready to analyse the
data under different assumptions. The presented formalism seems
perfect for this purpose. If the data by \citet{elshamouty13} are
 confirmed in future observations, the assumption by
\citet{pageetal2011} and \citet{shterninetal2011} that the cooling
is regulated by the effects of neutron superfluidity would remain
realistic and attractive explanation. The main indicator in favour
for this conclusion would be the observed value of the surface
temperature decline, $s_{\rm d}$; it has to be noticeably larger
than 0.1. If the data disfavour such large $s_{\rm d}$
\citep{posseltetal2013}, the theory can help imposing some
constraints on the properties of superfluidity in the stellar
core.

\section*{acknowledgements}

This work was partly supported by RFBR (grants 14-02-00868-a and
13-02-12017-ofi-M) and  RF Presidential Programme NSh-294.2014.2.

\appendix
\section{Calculation of cooling integrals}\label{s:app_im}
We deal with the family of integrals
\begin{equation}
  I_m(\tau)=\int_\tau^1 \frac{{\rm d}x}{x^m\,[1+\alpha (1-x)^2]},
\label{e:Im}
\end{equation}
with integer $m=0,1\dots$ and $\alpha=116\,\delta$. These
integrals satisfy useful recurrent relations
\begin{equation}
    (1+\alpha)\,I_m(\tau)=J_m(\tau)+2\alpha I_{m-1}(\tau)-\alpha I_{m-2}(\tau)
\label{e:recurr}
\end{equation}
with
\begin{eqnarray}
J_m(\tau)& = & \int_\tau^1 \frac{{\rm d}x}{x^m}=\frac{1}{m-1}
\,\left( \frac{1}{\tau^m} -1 \right),
\nonumber\\
I_1(\tau)&=& \frac{1}{2(1+\alpha)}\,\ln \left( \frac{1+\alpha
\,(1-\tau)^2}
  {\tau^2}\right) + \frac{\alpha}{1+\alpha}\,I_0(\tau) ,
\nonumber   \\
I_0(\tau)&=& \frac{1}{\sqrt{\alpha}}\;\mathrm{arctan}\,
\left(\sqrt{\alpha} \,(1-\tau)\right). \nonumber
\end{eqnarray}
 These relations allow one to calculate (\ref{e:Im}) at
 any $m$.



\end{document}